\documentclass[10pt,journal,cspaper,compsoc]{IEEEtran}
\usepackage{amsmath,amsthm}
\usepackage{authblk}
\usepackage{algorithmic}
\usepackage{algorithm}
\usepackage{stmaryrd}
\usepackage{pxfonts}
\usepackage{graphicx}
\usepackage{array}
\usepackage{booktabs}
\graphicspath{ {images/} }
\usepackage{tikz}
\usepackage{subcaption}
\newtheorem{definition}{Definition} 

\title{Impact zone analysis of p-cycle}

\author{Pallavi~Athe, Yatindra~Nath~Singh,~\IEEEmembership{Senior Member,~IEEE,}

\thanks{Pallavi Athe and Yatindra Nath Singh are with the Department of Electrical Engineering,
Indian Institute of Technology, Kanpur, India.}%
\thanks{E-mail:{\{apallavi, ynsingh}\}@iitk.ac.in}}

\begin{document}
\maketitle
\begin{abstract}
 Pre-configured cycle (p-Cycle) method has been studied in literature extensively for optical network protection. A large p-cycle has high capacity efficiency and can protect a large number of nodes against the single link failure scenarios. All the links protected by such a p-cycle lose protection when the p-cycle is consumed to restore traffic after a failure. As the probability of multiple link failure is high for a large network, it also means that with higher probability, on the second failure, protection may not be there for the failed link. Thus, if the number of links protected by a p-cycle is large, it makes the network unprotected with high probability on the advent of the second failure. In this paper, we study the impact zone due to a first link failure in the various configurations of the p-cycles. The study gives insight into how to choose the p-cycle configuration to reduce the impact zone while using minimum spare capacity. We propose few methods and compare them to show how the impact zone analysis can be used to improve the fault tolerance in an optical network.
\end {abstract}
\section{Introduction}
Survivability schemes are an integral part of optical network design. Survivability plays an essential role in an optical network by keeping it functional without interruption in the presence of failures. It involves protection and restoration. The protection is provisioning of redundant resources in the network, to be used during failures. Restoration is the process of establishing the failed connection again after failure, using redundant resources. An optical network which can survive a high number of simultaneous failures with minimum resources is desirable. The number of resources required increases with the increase in the failure protection capability of the network. Therefore, research efforts are mostly directed towards efficiently using the redundant resources to achieve resilience. Various network protection schemes have been proposed in the literature \cite{Ramaswami2010511} \cite{Simmons2014}  \cite{Zhou} \cite{Habib}. Ring based schemes have fast recovery (50-60 ms) whereas mesh based protection schemes are more capacity efficient. p-Cycles possess both, the speed of ring and efficiency of mesh protection. Grover and Stamatelakis proposed the concept of p-cycles in 1998 \cite{Grover1}. There has been significant amount of research done on p-cycles since then \cite{Asthana} \cite{Grover2006_chapter} \cite{Pathp} \cite{Wang2016} \cite{NEPC_Grover} \cite{Ju:16} \cite{EON}. p-Cycles are pre cross-connected ring-like structures formed from the spare capacity of the network. One of the main difference which distinguishes p-cycles from the ring is their ability to restore on-cycle as well as the straddling link failures. On-cycle links of a p-cycle are the links which form the p-cycle. Straddling links are the chords with there end nodes on the p-cycle.\\
 Two-link failure protection schemes are desirable due to a significant probability of frequent two-link failures. Maximizing the number of deployed p-cycles improves dual failure restorability of a network \cite{Schupke2}. Dual failure restorability also improves on minimizing the maximum coverage of working capacity by the p-cycles in a network \cite{Schu1}. In \cite{Akpuh}, Akpuh \emph{et al}  introduced a method to design a network with specified dual failure restorability. In \cite{Mltipleqos}, dual failure protection is provided to the links of the network depending on the quality of service of the link. Apart from these, there are some studies with the deterministic approach. Feng \emph{et-al} \cite{Feng} proposed a method in which each working link in a network is protected by two p-cycles with link disjoint segments.  The concept of a pre-configured ball (p-ball) is proposed by Chen Ma \emph{et al} \cite{p-ball}. p-Ball is a subgraph with ring set which uses minimum backup links for dual link failure protection. Wang \emph{et al} \cite{Wang} introduced MFS (multi failure survivability scheme) to provide multi-failure survivability based on p-cycles. Huang \emph{et al} \cite{Polyhed} proposed pre-configured polyhedron (PCP) based protection structure to protect both simultaneous and sequential random link failures with improved spare resource efficiency. The scheme requires searching backup paths within the PCP after any failure to restore the traffic.\\
The studies in the literature focus mainly on the reduction of the spare capacity. There is no study which quantifies the effect on working capacity after the restoration of any fault. In this paper, we analyze the impact of a link failure on the protection of working capacity of the optical network protected by p-cycles. We define impact zone which mathematically represents the effect of the link failure on protection capacity of the network. Various methods are proposed to reduce the impact zone, to increase the double link failure tolerance of the network. In Section 2, we define and illustrate mathematical formulation of the impact zone. In Section 3, integer linear program (ILP) for optimizing spare capacity for single link failure has been formulated.  We also introduced and discussed limiting constraints which can reduce impact zone in Section 3. In Section 4, simulation results have been discussed, and Section 5 concludes the work presented in this paper.  
\section{Impact zone}
\begin{figure}
       \includegraphics[width=8cm]{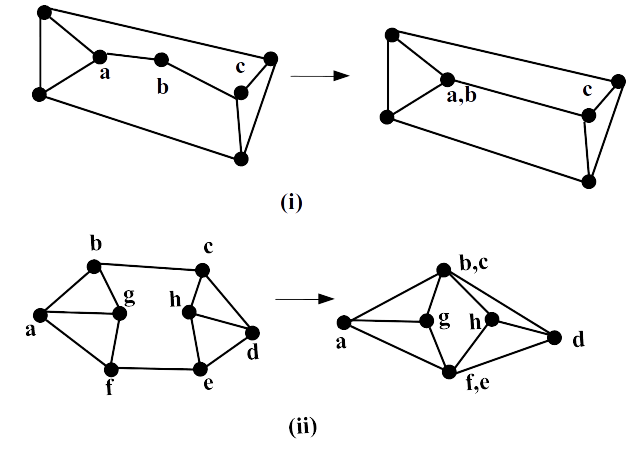}
       \centering
       \caption{ Illustration to modify a graph to a 3-connected graph}
       \label{merge}
    \end{figure}

 Protection capacity of a p-cycle is shared among a set of links which can be protected by it. Therefore, full protection of only one of these links can be ensured. The other links may lose full or partial protection after the traffic disrupted by the first failure is restored. Following links will lose protection on the restoration of traffic on the failed link:
 \begin{enumerate}
 \item The links which share the same p-cycle with the failed link for protection.
 \item The links which are protected by the p-cycles having the failed link as the on-cycle link.
 \end{enumerate}
 We can define impact zone of a link as the working capacity which loose protection due to the utilization of p-cycle to protect the failed link. The impact zone has been formulated to quantify the double fault tolerance for a network which is deterministically 100\% protected against single link failure. Therefore,  We consider networks that have 100\% single link failure protection.\\
 In this paper, the considered optical networks have been represented by bidirectional graphs. For a network to survive any two link failures, it should be 3-connected \cite{Polyhed} \cite{pkecks} \cite{3-connected1}. A 3-connected graph is a graph that has three links disjoint path between any pair of nodes \cite{nla.cat-vn2190177}. Since we are investigating double fault tolerance, We have used 3-connected graphs for simulation and analysis. Non 3-connected graphs have been modified to 3-connected graphs by applying merge operation on them. A graph can be modified to a 3-connected graph by identifying the pair of nodes having less than three disjoint paths and performing following merge operations on them.
   \begin{enumerate}
   \item A node in a graph with degree two can be merged to a neighboring node. As an example, in Fig. \ref{merge} (i), node $b$ with degree two is merged with node $a$. 
   \item Node-pairs in a graph having less than three disjoint paths can be merged as illustrated in Fig. \ref{merge} (ii) to modify the graph to a 3-connected graph.
     \end{enumerate}     
Notations and variables used for the optical network are as follows:\\
S=set of spans.\\
P=set of simple cycles.\\
$w_{i}$=working capacity on span  $i$.\\
$n_{p}$ = number of copies of p-cycle  $p$.\\
$n_{i,p}$=number of copies of p-cycle $p$ required to protect link $i$\\ 
$c_{k}$ =Overall protection capacity available for link $k$.\\
We define $x_{i,p}$ as follows:
 \[x_{i,p} =
 \begin{cases}
   1 & \quad \text {if span \textit{i}  is on-cycle span of p-cycle \textit{p},} \\
   2 & \quad \text{if span \textit{i}  is straddling link of p-cycle \textit{p}, and} \\
   0 & \quad \text{otherwise.}
   \end{cases}
 \]
  We define an indicator function $I_{A}(x)$ as follows:
  \[I_{A}(x)=
  \begin{cases}
    |x| & \quad \text{if $x<0$,}\\
    0 & \quad \text{if $x\geq0$}.\\
    \end{cases}
  \] 
 The impact on the protection capacity depends on whether the failed link and the protected links are on-cycle or straddling links on the p-cycle/p-cycles affected by the failed link. The impact for these cases is formulated below.
 \begin{enumerate}
\item The failed link $i$ and other protected link $k$ both are on-cycle on p-cycle $p$. When the link $i$ fails, the p-cycle is broken and the whole of the protection given by p-cycle $p$ will be lost. The protection capacity provided by the p-cycle $p$ to link $k$ which is lost due to the failure of the link $i$ is given as:
 \begin{equation}\label{I1}
 I_{i,k,p}^{'}=n_{k,p} 	
 \end{equation} 
\item The failed link $i$ is on-cycle and other protected link $k$ is straddling on p-cycle $p$. Impact due to the failure of link $i$ on protection capacity of link $k$ (loss of protection) is given as:
 \begin{equation}\label{I2}
 I_{i,k,p}^{'}= I_{A}((n_{p}-n_{i,p})-2n_{k,p}).	
 \end{equation} 
\item If link $i$ is straddling link on p-cycle $p$, the other protected link $k$ can be on-cycle or straddling on p-cycle $p$, and $n_{i,p}>0$, the loss of protection for link $k$ is 
 \begin{equation}\label{I3}
 I_{i,k,p}^{'}= I_{A}(x_{k,p}[(n_{p}-n_{i,p})-n_{k,p}]).	
 \end{equation}
 \end{enumerate}
 Loss of protection capacity of link $k$ due to failure of link $i$ is
 \begin{equation}\label{IZk}
 I_{i,k}^{'}=\sum_{p\in P}I_{i,k,p}^{'}.
 \end{equation}
 The capacity available to protect link $k$ after failure of link $i$ is thus $c_k
-I_{i,k}^{'} $. Since we have considered deterministic single link protection $c_k \geq w_{k}$. If $c_{k}-I_{i,k}^{'}\geq w_{k}$ then, link $i$ gets full protection even after failure of link $i$. But, if $c_{k}-I_{i,k}^{'}< w_{k}$ then, $w_{k}-(c_{k}-I_{i,k}^{'})$ capacity of link $k$ remains unprotected in case of failure of link $i$. Thus, the impact zone $I_{i}$ of link $i$, is given as.
 \begin{equation}\label{IZ}
 I_{i}=\sum_{\substack{k\in S\\I_{i,k}^{'}\neq 0}}I_{A}(c_{k}-I_{i,k}^{'}- w_{k}).
 \end{equation}
\subsection{Example to calculate impact zone}
\begin{figure}
       \includegraphics[width=7cm]{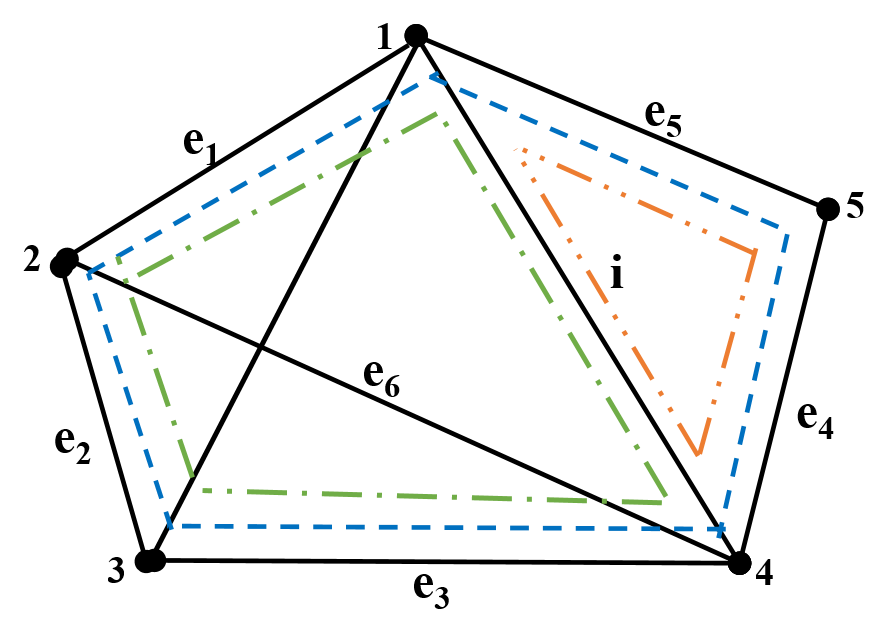}
       \centering
       \caption{ Example of a network for calculating Impact zone of link $i$.}
       \label{fig:Imz }
\end{figure}

Consider the network shown in Fig \ref{fig:Imz }. we calculate impact zone for link $i$ in Fig. \ref{fig:Imz }. Let p(1-2-3-4-5-1), q(1-2-3-4-1) and r(1-4-5-1) are the p-cycles associated with link $i$ which protect links ($e_{1}$, $e_{2}$,$i$),($e_{3}$, $e_{6}$,$i$) and ($e_{4}$,$e_{5}$) respectively. When link $i$ fail, p-cycle $p$ and $q$ will be used to protect the link $i$. Links $e_{1}$-$e_{6}$ will lose protection fully or partially depending on the number of copies of p-cycle $p$ used to restore link $i$. The remaining copies of p-cycle $p$ will be $n_{p}-n_{i,p}$. Protection lost by $e_{1}$ and $e_{2}$ will be $n_{e_{1},p}+n_{e_{2},p}-2(n_{p}-n_{i,p})$. Link $e_{3}$ is on-cycle on p-cycle $q$ therefore, $n_{e_{3},q}$ capacity of $e_{3}$ will lose protection. Link $e_{6}$ is a straddling link on p-cycle $q$. p-Cycle $q$ protects $2n_{e_{6},q}$ working capacity of link $e_{6}$. The protection lost by $e_{6}$ will be $2n_{e_{6},q}-(n_{q}-n_{i,q})$. Link $e_{4}$ and $e_{5}$ are on-cycle on p-cycle $r$ and, lose protection by $n_{e_{4},r}$ and $n_{e_{5},r}$ respectively. The impact zone of link $i$ will be sum of protection capacity lost by links $e_{1}$-$e_{6}$ is given by 
\begin{multline}\label{IZi}
I_{i}=n_{e_{1},p}+n_{e_{2},p}-2(n_{p}-n_{i,p})+n_{e_{3},q}+2n_{e_{6},q}-(n_{q}-n_{i,q})\\+n_{e_{4},r}+n_{e_{5},r}
\end{multline}
\section{Improving double fault tolerance}
An optical network protected for single link failure will have some inherent double link failure tolerance. Reducing impact zone of the network results in the increase in its double fault tolerance capability. In the following subsection, we state an ILP for single link failure protection \cite{Grover1} where additional constraints have been incorporated to reduce its impact zone thereby increasing its tolerance to double link failure, i.e. failure of one more link after the first one. 
\subsection{ILP for single link protection}
In this section, we formulate and describe the ILP for single link failure protection (SLP) in an optical network. The ILP minimizes the total spare capacity required to protect the optical network from single link failure.\\
\textbf{Notations}\\
In addition to the notation defined in section 2, we define following notation to formulate the ILP for single link failure protection of optical network.\\
$c_{k}$=cost of unit capacity of span \textit{k}.\\
$s_{k}$=spare capacity on span \textit{k}.\\
$a_{i,p}$ is a binary variable associated with link $i$ and p-cycle $p$.
$M$ is an arbitrary large positive number.
\[\delta_{i,p}=
\begin{cases}
  1 & \quad \text{if span \textit{i} is on-cycle on p-cycle \textit{p}},\\
  0 & \quad \text{otherwise}.\\
  \end{cases}
\]

\textbf{Minimize:} 
\begin{equation}\label{eq:1}
\sum_i\in S c_{i}s_{i} 
\end{equation}
\textbf{Subject to:}
\begin{equation}\label{eq:2}
 \sum_{\forall p\in P}x_{i,p}.n_{i,p} \geq w_{i}	\qquad \forall i \in S
\end{equation}
\begin{equation}\label{eq:3}
n_{p}\geq n_{i,p}\delta_{i,p} \quad \forall p\in P; i\in S
\end{equation}
\begin{equation}\label{eq:4}
s_{i}\geq\sum_{p \in P} n_{i}\delta_{i,p}\qquad \forall i\in S
\end{equation}
\begin{equation}\label{eq:5}
n_{p}\geq 0,n_{i,p}\geq 0 ; \quad \forall i \in S ;\forall p \in P
\end{equation}
 We need to solve the ILP to find minimum spare capacity to provide 100\% single link failure protection. The objective function given by equation \ref{eq:1} minimizes the total spare capacity on each link of the optical network. Constraint \ref{eq:2} ensures that the number of copies of the p-cycles are sufficient to protect the working capacity of the links. Constraint \ref{eq:3} selects the number of copies of p-cycle required to protect the maximum working capacity link among all the links protected by the p-cycle. Constraint \ref{eq:4} ensures that spare capacity on each link is sufficient to form the p-cycles. It should be noted that variable $n_{i,p}$ is required to calculate the impact zone. Therefore, additional variable $n_{i,p}$ and constraint \ref{eq:3} are required in the formulated ILP.
\subsection{Minimizing impact zone}
Solution of ILP given by eqn. \ref{eq:1} to \ref{eq:5} for a given optical network provides required optimum spare capacity and p-cycles to protect the network from single link failure. The impact zone for the links in the network is computed using the variables obtained by solving the ILP (eqn. \ref{eq:1} to \ref{eq:5}) in eqn. \ref{I1} to \ref{IZk}. A lower value of the impact zone indicates better double link failure tolerance.\\
We can limit the value of impact zone by adding limiting-constraints to the ILP of single link protection (eqns \ref{eq:1} to \ref{eq:5}). The limiting-constraints control the variables involved in the impact zone. This approach reduces the impact zone by reducing the sharing of p-cycles among links.
\subsubsection{Limiting-constraint 1}
 Constraints \ref{eq:2} to \ref{eq:5} ensures 100\% single link failure protection in a network. We add limiting-constraint 1 (L1) given by eqn. \ref{c1} to reduce the value of impact zone. The left hand side of L1 contains all the variables associated with link $i$ which can affect the protection capacity of other links. 
\begin{equation}\label{c1}
\begin{split}
	\sum_{p \in P}\sum_{ \substack{j\in S\\j\neq i\\x_{i,p}=1\\x_{j,p}=1}}n_{j,p}+\sum_{ p \in P}\sum_{\substack{ j\in S\\j\neq i\\x_{i,p}=1\\ x_{j,p}= 2}}[n_{j,p}-(n_{p}-n_{i,p})]+\sum_{ p \in P}\sum_{\substack{ j\in S\\j\neq i\\x_{i,p}=2\\ x_{j,p}= 2}}2[n_{j,p}-(n_{p}-n_{i,p})] \\
	 +\sum_{ p \in P}\sum_{\substack{ j\in S\\j\neq i\\x_{i,p}=2\\ x_{j,p}= 1}}[n_{j,p}-(n_{p}-n_{i,p})]\leq K; \qquad 	\forall i \in S.\qquad \qquad \qquad
	\end{split}
\end{equation} 
Here, $K$ is a constant. For a sufficiently large value of $K$ the solution of ILP will converge to the solution of pure single link protection ILP. Therefore, impact zone is also same for a large value of $K$, as for the ILP without any limiting constraint. Impact zone decreases as we reduce the value of $K$.  \\
\subsubsection{Limiting-constraint 2}
Another possible limiting-constraint 2 (L2) is given by eqn. \ref{c2}. It is applied to the ILP of single link protection (eqn. \ref{eq:1} to \ref{eq:5}) to reduce the sharing of p-cycles among links.
\begin{equation}\label{c2}
\sum_{ \substack{\forall i\in S\\x_{i,p} \ne 0}}n_{i,p} \le n_{p} \quad \quad \forall p\in P.
\end{equation}
Eqn. \ref{eq:3} along with eqn. \ref{c2} reduces the number of links which share a p-cycle. Thus the spare capacity required will increase and impact zone will decrease.\\
\subsubsection{Limiting-constraint 3}
We can also add the another additional limiting-constraints 3 (L3), represented by eqn. \ref{c3} to \ref{c5} to the ILP eqn. \ref{eq:1} to \ref{eq:5}. L3 ensures that each p-cycle protects only one link. Thus, there is no sharing of p-cycles among links. We use a variable $a_{i,p}$ with the conditions such that it can be either zero or one. $M$ is a sufficiently large number ($M > \sum_{ \substack{\forall i\in S}}w_{i}$).\\
\begin{equation}\label{c3}
a_{i,p}\geq \frac{n_{i,p}}{M} \quad \forall i \in S. 
\end{equation} 
\begin{equation}\label{c4}
a_{i,p}\leq 1 \quad \forall i \in S. 
\end{equation}
\begin{equation}\label{c5}
\sum_{ \substack{\forall i\in S\\x_{i,p} \ne 0}}a_{i,p} \le 1 \quad \quad \forall p\in P.
\end{equation}
Constraint \ref{c3} and \ref{c4} ensure that the variable $a_{i,p}$ can take binary value (0 or 1). Constraint \ref{c4} ensures that each p-cycle protects only one link.
\section{Results and discussion}
\begin{figure}
       \includegraphics[width=\linewidth]{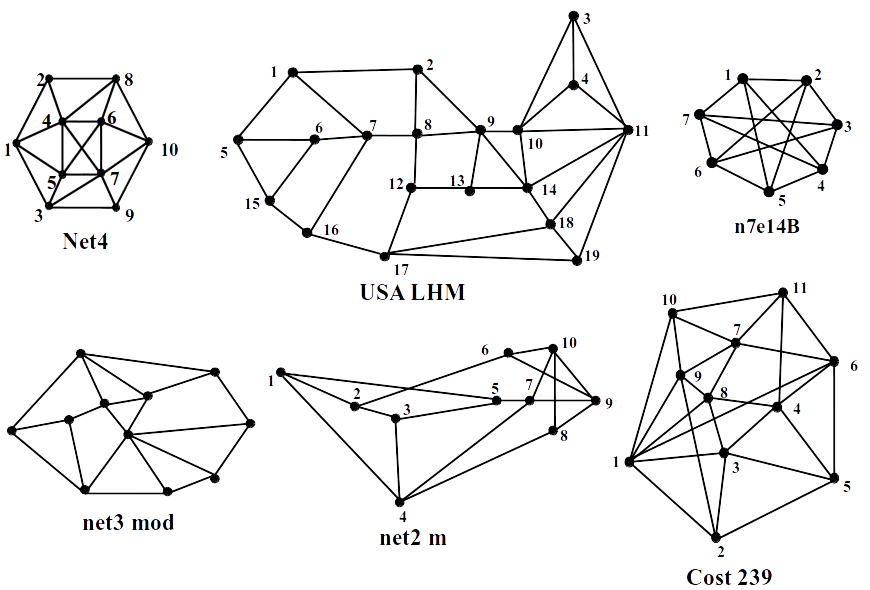}
       \centering
       \caption{Network graphs used for simulation.}
       \label{NW}
\end{figure}

In this section, we analyze the effect of limiting-constraints L1, L2 and L3 in terms of spare capacity and impact zone. Computation is performed for the networks shown in Fig. \ref{NW}. The networks net2m, Cost239 (Pan European network), net3mod, Net4, USA LHM (USA long Haul) are taken from \cite{cost239net2}, \cite{Net3} and \cite{USAlonghaul}. The networks net2 (also known as NSFNET), USA long Haul and net3 are modified to 3-connected graphs by performing the merge operation in the network as explained in section 2. Scilab 5.5  has been used for ILP formulation. ILOG CPLEX 9 has been used to solve the ILP on AMD Opteron(tm) 1.8 GHz CPU. One of the measure of performance of p-cycle based protection is Spare capacity efficiency (SE). It is defined as follows.
\begin{definition}\label{Spare capacity efficiency}
 Spare capacity efficiency (SE) is the ratio of spare capacity required for protection and the working capacity in the network. It is given by 
  \begin{equation}\label{SE} 
    SE = \frac{\sum\limits_{\forall i\in S} s_{i}}{\sum\limits_{\forall i\in S}w_{i}}
    \end{equation} 
     where $s_{i} $ and $ w_{i}$ are spare and working capacity on link $i$ respectively.
\end{definition}
Low value of SE is desirable in any protection scheme.
\subsection{Performance of L1, L2 and L3}
\begin{table*}
        \centering
        \caption{Simulation Results}
        \label{IZtable}
    \begin{tabular}{c|cc|cc|cc|cc|cc}\toprule
     & \multicolumn{2}{c|}{SLP}&\multicolumn{2}{c|}{L1(K=0)} & \multicolumn{2}{c|}{L1(K=max)} & \multicolumn{2}{c|}{L2}& \multicolumn{2}{c}{L3}\\
     \cmidrule{2-11} 
 Network & SE & Impact  & SE & Impact &  SE & Impact &  SE & Impact &  SE & Impact \\
  &   &   Zone &   & Zone &   & Zone &    & Zone &    & Zone \\
     \midrule 
   net3mod & 0.59 & 155.5 & 2.79 & 18.45 &  0.59 & 134.3 &  2.85 & 0 &  2.85 & 0 \\
   net2m & 0.8 & 106 & 3.62 & 21.5 &  0.8 & 80.1 &  3.50 & 1.2 &  3.59 & 0 \\
   n7e14B & 0.54 & 39 & 2.18 & 7.9 &  0.54 & 32 &  2.29 & 0 &  2.29 & 0 \\
   Net4 & 0.47 & 106.8 & 2.06 & 15.5 &  0.47 & 88.7 &  2.34 & 0.2 &  2.36 & 0 \\
   USA LHM & 0.71 & 532 & 3.36 & 51.7 &  0.71 & 414 &  3.22 & 3.3 &  3.24 & 0 \\
   cost239 & 0.43 & 106 & 1.90 & 17.8 &  0.43 & 99.9 &  2.31 & 0.08 &  2.31 & 0 \\
          \bottomrule
         \end{tabular}
           \end{table*}

\begin{figure*}
  \centering
  \begin{subfigure}[b]{0.45\textwidth}
    \includegraphics[width=8cm]{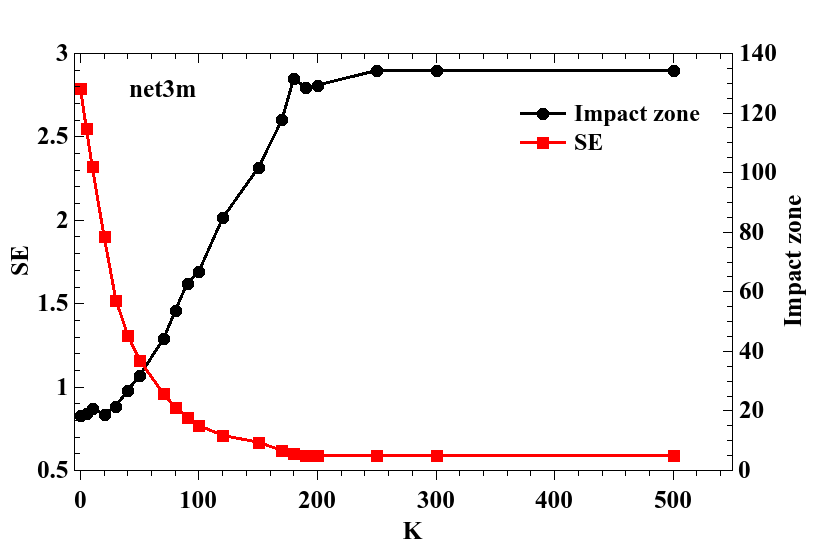}
    \subcaption{net3m}
  \end{subfigure}
  \begin{subfigure}[b]{0.45\textwidth}
    \includegraphics[width=8cm]{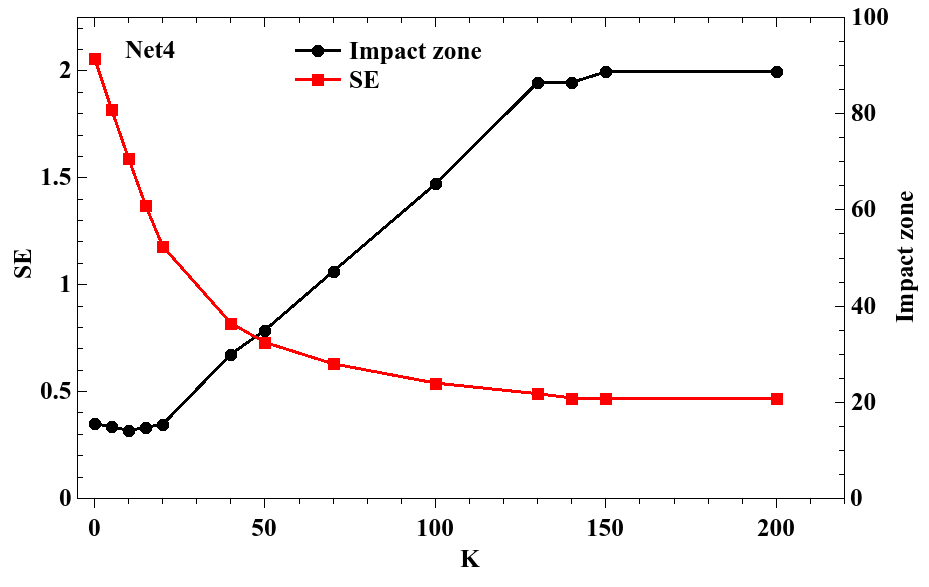}
    \subcaption{net4}
  \end{subfigure}
  \begin{subfigure}[b]{0.45\textwidth}
    \includegraphics[width=8cm]{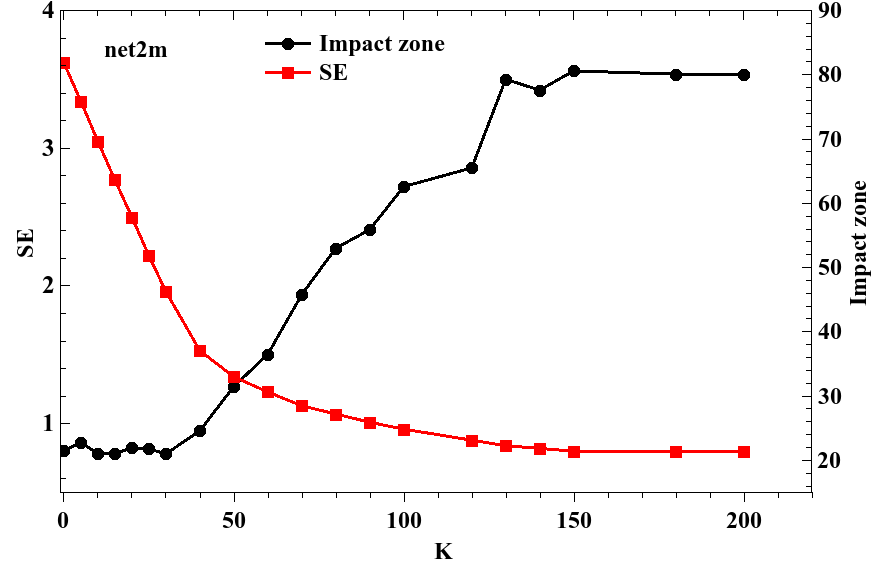}
    \subcaption{net2m}
  \end{subfigure} 
    \begin{subfigure}[b]{0.45\textwidth}
      \includegraphics[width=8cm]{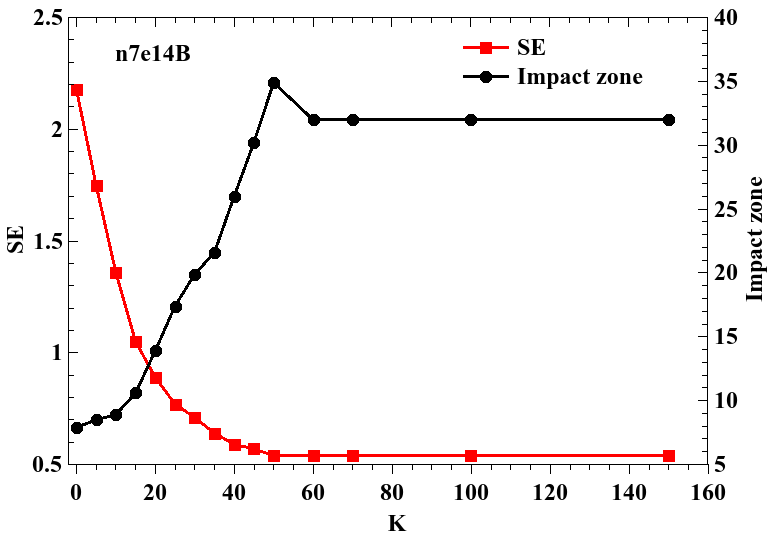}
      \subcaption{n7e14B.}
    \end{subfigure} 
    \begin{subfigure}[b]{0.45\textwidth}
      \includegraphics[width=8cm]{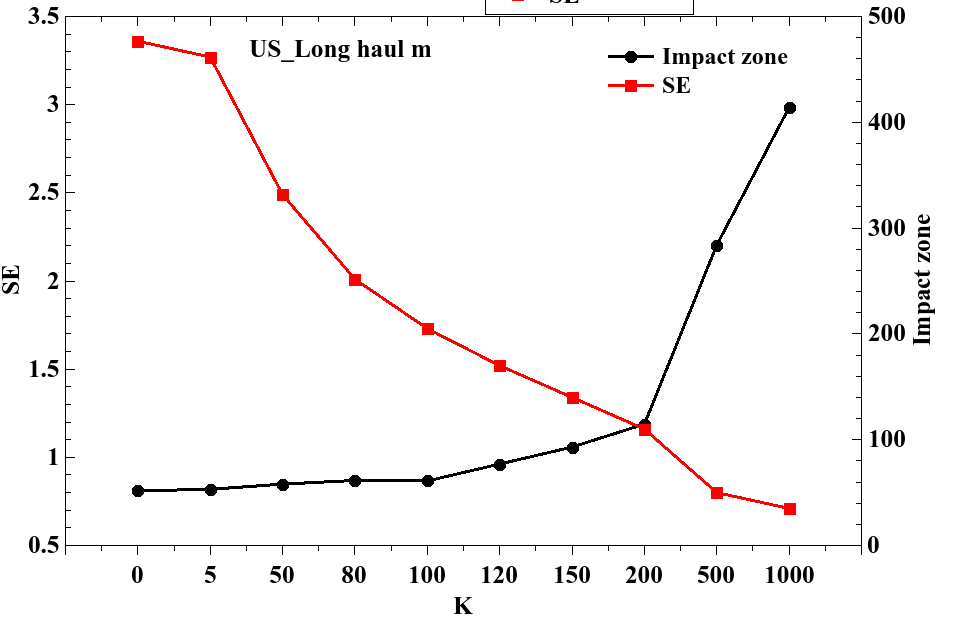}
      \subcaption{USA Long Haul m}
    \end{subfigure}
        \begin{subfigure}[b]{0.45\textwidth}
          \includegraphics[width=8cm]{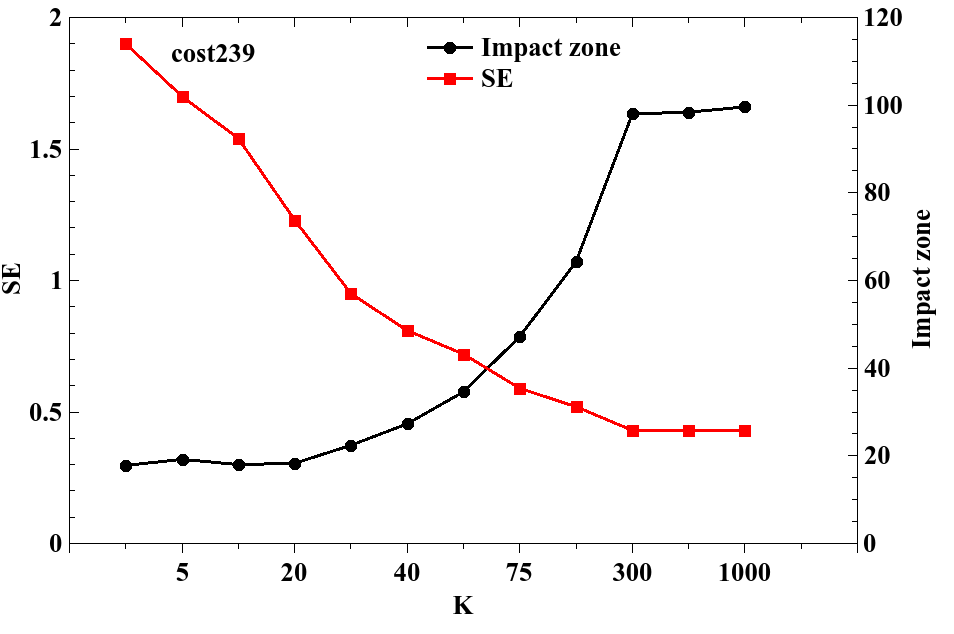}
          \subcaption{cost239}
        \end{subfigure}    
    
 \caption{Effect of varying $K$ on impact zone and SE. }
 \label{K var}   
\end{figure*}

Single link protection ILP eqn. (\ref{eq:1} to \ref{eq:5}) with L1 (eqn. \ref{c1}) is solved for different values of $K$. L1 given by eqn. \ref{c1} controls the variable associated with a link which can affect the protection capacity of other links. The value of impact zone increases while, SE decreases with increase in $K$ (Fig.\ref{K var}). A lower value of impact zone implies greater fault tolerance capability which in turn requires higher spare capacity. Hence there is a trade-off between SE and impact zone. The value of $K$ can be chosen to obtain the desired value of SE and impact zone. It should be noted that the value of impact zone does not reduce to zero for $K=0$. $K$ does not represent the true value of impact zone. For $K=0$, the expression on the left-hand side of eqn. \ref{c1} reduces to zero. The expression for impact zone may not contain all the variables present in L1, and therefore it has nonzero value. \\
 
\begin{figure*}
  \centering
  \begin{subfigure}[b]{0.45\textwidth}
    \includegraphics[width=7cm]{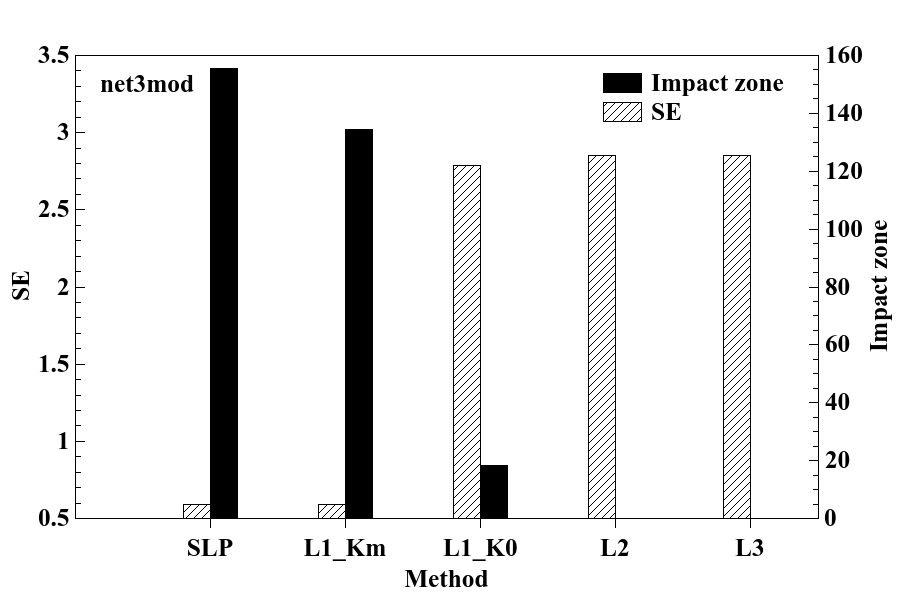}
    \subcaption{net3m}
  \end{subfigure}
  \begin{subfigure}[b]{0.45\textwidth}
    \includegraphics[width=7cm]{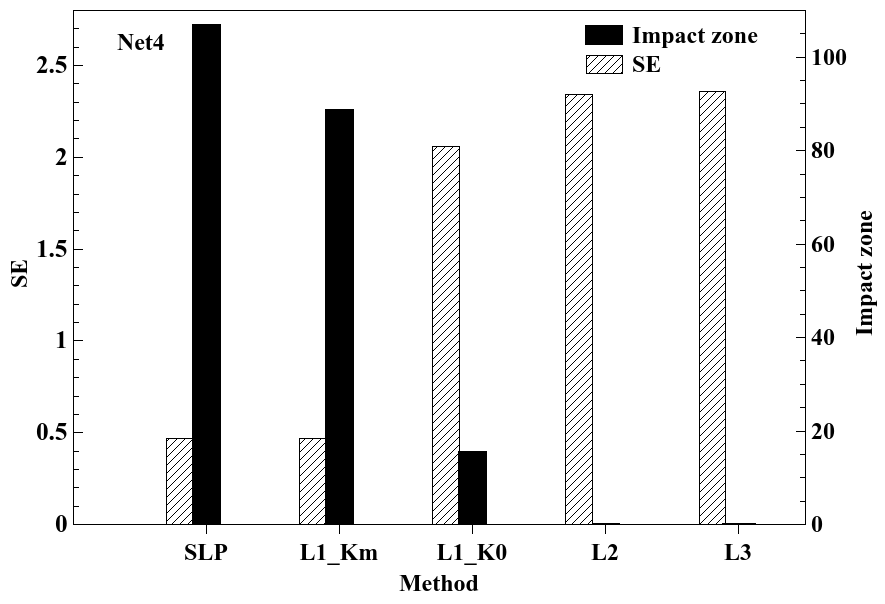}
    \subcaption{net4}
  \end{subfigure}
  \begin{subfigure}[b]{0.45\textwidth}
    \includegraphics[width=7cm]{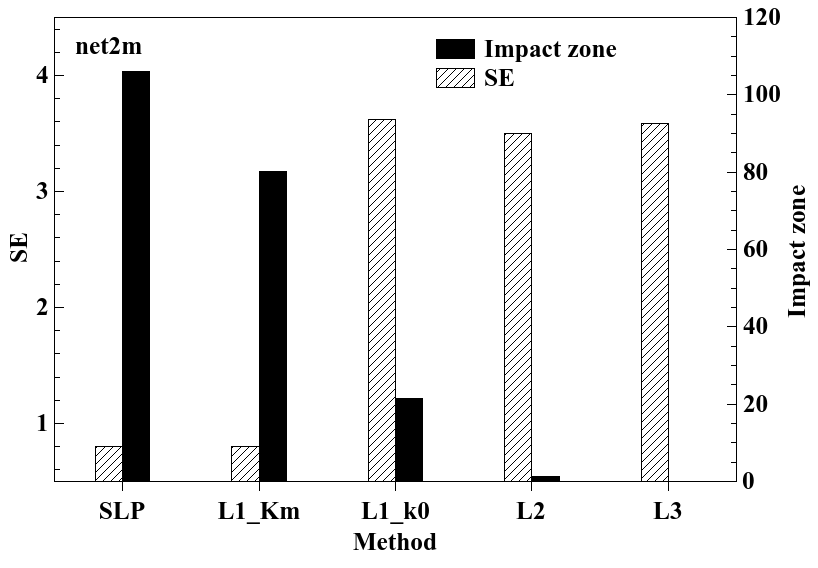}
    \subcaption{net2m}
  \end{subfigure} 
    \begin{subfigure}[b]{0.45\textwidth}
      \includegraphics[width=7cm]{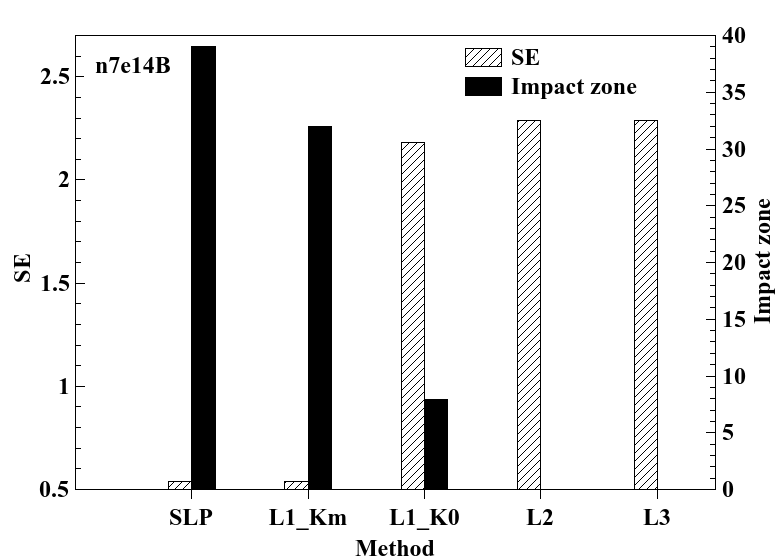}
      \subcaption{n7e14B.}
    \end{subfigure} 
    \begin{subfigure}[b]{0.45\textwidth}
      \includegraphics[width=7cm]{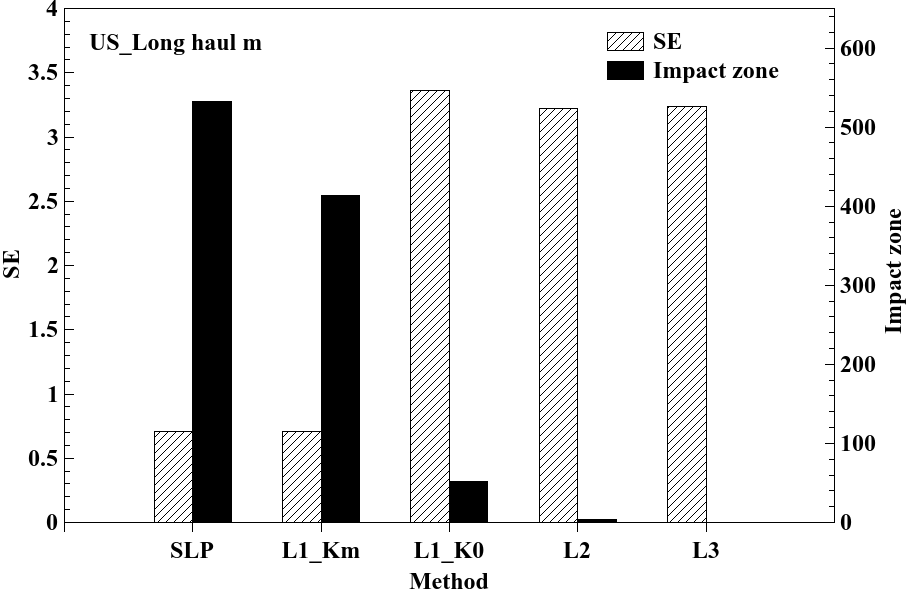}
      \subcaption{USA Long Haul m}
    \end{subfigure}
        \begin{subfigure}[b]{0.45\textwidth}
          \includegraphics[width=7cm]{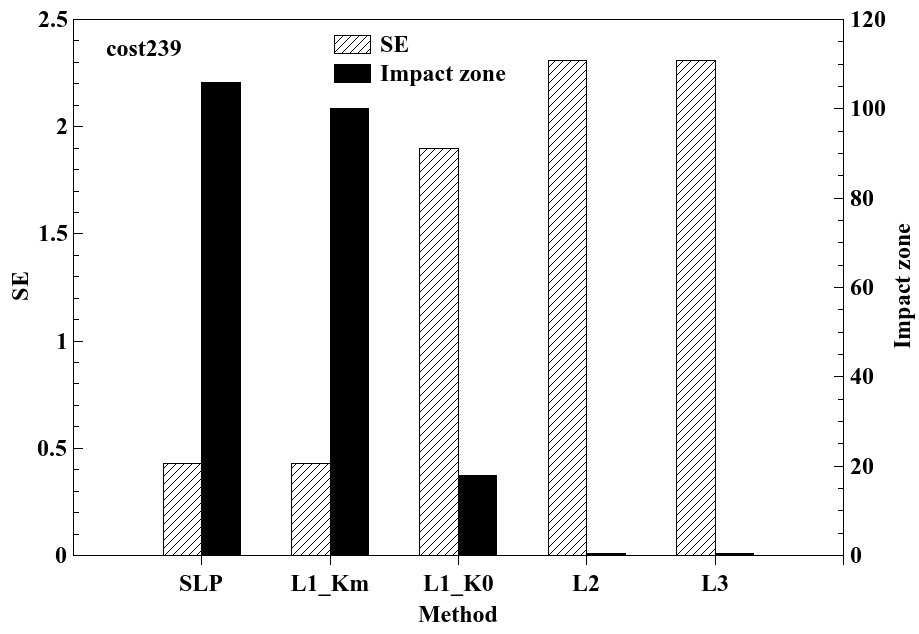}
          \subcaption{cost239}
        \end{subfigure}    
    
 \caption{Effect of different limiting constraints on Impact zone and $SE$ }  
 \label{Limitvar} 
\end{figure*}

Various protection methods described in section 3, SLP, SLP with L1, SLP with L2, SLP with L3 have been compared in Fig. \ref{Limitvar} and table \ref{IZtable}. In Fig. \ref{Limitvar},  \texttt{$L1\_K0$} represents the result with L1 and $K=0$ while, \texttt{$L1\_Km$} represents the result with L1 and value of $K$ for which SE and impact zone for the given network saturates. We observed that impact zone is lowest with L2 and L3 constraint in all the networks. The value of SE is higher for \texttt{$L1\_Km$}, L2, and L3 as compared to SLP and \texttt{$L1\_K0$}. L2 and L3 are comparable and give better double link failure tolerance than SLP, \texttt{$L1\_K0$} and \texttt{$L1\_Km$}. L1 with a suitable value of $K$, L2 and L3 can be useful for the networks that need higher double fault tolerance.\\

\section{Conclusion}
The impact zone is a measure of protection lost by a network after a failure and restoration of the working capacity passing through the failed link. Double link failure protection capability of a network can be improved by various techniques to reduce the impact zone. Different limiting constraints can be added to single fault protection ILP to minimize the impact zone. The value of impact zone and SE depends on the value of $K$ in the limiting constraint L1. L1 is able to provide some control over the value of impact zone and SE. Limiting constraint L1 can be used to tune to the desired value of impact zone and SE but, L1 does not provide the lowest possible value of impact zone. L2 and L3 reduce the impact zone by limiting the sharing of p-cycles among links. We observed that L2 and L3 give better double fault tolerance as compared to L1 and SLP in terms of impact zone while sacrificing SE.\\
There is a trade-off between impact zone and SE. Both SE and impact zone should be considered while designing a survivable optical network. Choice of the method might depend on the design requirement of the network protection.

\bibliographystyle{IEEEtran}
\bibliography{ref.bib}
\end{document}